\documentclass[conference]{IEEEtran}\IEEEoverridecommandlockouts
\usepackage{etoolbox}
\makeatletter
\patchcmd{\@makecaption}
  {\scshape}
  {}
  {}
  {}
\makeatletter
\patchcmd{\@makecaption}
  {\\}
  {.\ }
  {}
  {}
\makeatother
 \usepackage{amsmath,amssymb}
 \usepackage{subfigure}
 \usepackage{graphicx,graphics,color,psfrag}
 \usepackage{cite,balance}
 \usepackage{caption}
 \allowdisplaybreaks
 \usepackage{algorithm}
 \usepackage{accents}
 \usepackage{amsthm}
 \usepackage{bm}
 \usepackage{algorithmic}
 \usepackage[english]{babel}
 \usepackage{multirow}
 \usepackage{enumerate}
 \usepackage{cases}
 \usepackage{stfloats}
 \usepackage{dsfont}
 \usepackage{color,soul}
 \usepackage{amsfonts}
 \usepackage{cite,graphicx,amsmath,amssymb}
 \usepackage{subfigure}
 \usepackage{fancyhdr}
 \usepackage{hhline}
 \usepackage{graphicx,graphics}
 \usepackage{array,color}
 \usepackage{amsmath}
\usepackage{float}
\usepackage{amssymb}
\usepackage{amsmath}
\usepackage{amsthm}
\usepackage{amsfonts}
\usepackage{graphicx}
\usepackage{algorithm}
\usepackage{algorithmic}
\usepackage{epstopdf}
\usepackage{cite}
\usepackage{amsmath,bm}
\usepackage{subfigure}
\usepackage{graphicx}
\usepackage{color}
\usepackage{graphicx}
\usepackage{calc}
\usepackage{caption}

\newtheorem{theorem}{\textbf{Theorem}}

\newtheorem{lemma}{\textbf{Lemma}}

\newtheorem{Prob}{\textbf{Problem}}

\begin{document}
\title{Modeling and Trade-off for Mobile Communication, Computing and Caching Networks}
\author{Yaping Sun, Zhiyong Chen, Meixia Tao and Hui Liu\\
Department of Electronic Engineering, Shanghai Jiao Tong University,  Shanghai,  P. R. China\\
Email: \{yapingsun, zhiyongchen, mxtao, huiliu\}@sjtu.edu.cn}
\maketitle
\begin{abstract}
Computation task service delivery in a computing-enabled and caching-aided multi-user mobile edge computing (MEC) system is studied in this paper, where a MEC server can deliver the input or output datas of tasks to mobile devices over a wireless multicast channel. The computing-enabled and caching-aided mobile devices are able to store the input or output datas of some tasks, and also compute some tasks locally, reducing the wireless bandwidth consumption. The corresponding framework of this system is established, and under the latency constraint, we jointly optimize the caching and computing policy at mobile devices to minimize the required transmission bandwidth. The joint policy optimization problem is shown to be NP-hard, and based on equivalent transformation and exact penalization of the problem, a stationary point is obtained via concave convex procedure (CCCP). 
Moreover, in a symmetric scenario, gains offered by this approach are derived to analytically understand the influences of caching and computing resources at mobile devices, multicast transmission, the number of mobile devices, as well as the number of tasks on the transmission bandwidth. Our results indicate that exploiting the computing and caching resources at mobile devices can provide significant bandwidth savings.
\end{abstract}
\section{Introduction}\label{I}
\begin{figure}[t]
\begin{center}
 \includegraphics[width=8cm]{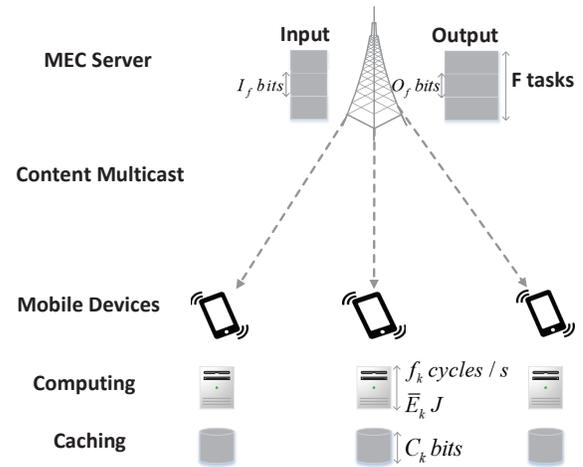}
\end{center}
 \caption{\small{A MEC system consisting of one MEC server and $K$ computing-enabled and caching-aided mobile devices.}
 }
\label{model}
\end{figure}

Computation task service delivery in a computing-enabled and caching-aided multi-user mobile edge computing (MEC) system is studied in this paper, as illustrated in Fig.~\ref{model}. Each computation task $f\in\mathcal{F} \triangleq \{1,2,\cdots,F\}$ is characterized with  input size $I_f$ (\emph{in bits}), computation load $w_f$ (\emph{in cycles/bit}) and an output size $O_f$ (\emph{in bits}). A set $\mathcal{K} \triangleq \{1,2,\cdots,K\}$ of $K$ mobile devices is connected to a MEC server (e.g., base station (BS)) over a wireless multicast channel. Each mobile device $k$ is endowed with limited computing (rate $f_k$ (\emph{in cycles/s}), average energy $\bar{E}_k$ (\emph{in J})) and caching $C_k$ (\emph{in bits}) resources, so as to store input or output datas of some tasks, as well as compute some tasks locally. Based on the caching and computing decision at the mobile devices, each task request at a mobile device can be served via local output or input caching, local computing or MEC computing, each of which yields a different bandwidth requirement.

\textbf{\textit{Motivating Example.}} One example of the above system is mobile virtual reality (VR) delivery, which generally leads to ultra-high transmission rate requirement (on the order of \emph{G bits/s}), deemed as a first killer application for $5G$ wireless network \cite{E}. In the VR framework, the projection component can be computed at the MEC server or at the mobile VR devices due to its low computational complexity \cite{Sunvr}. Specifically, compared with computing at the MEC server, computing at the mobile VR device can reduce at least half of the traffic load on the wireless link, since the data size of the output (i.e., 3D FOV) is at least twice larger than that of the input (i.e., 2D FOV). However, computing at the mobile VR device may incur longer latency, since the computing capability of the mobile VR device is generally weaker than that of the MEC server. Thus, \textit{the computing policy}, i.e., computing the projection either at the MEC server or at the mobile VR device, requires careful design. In addition, caching capability of each mobile VR device can be utilized to store some input or output datas for future requests. Specifically,  compared with caching the input data of some task, 
caching the output data can help reduce both latency and energy consumption, since the VR video request can be served directly from local caching and with no need of computing. However, output caching consumes larger caching resource at the mobile VR device, since output data size is at least twice larger than input data size. Thus, \textit{the caching policy}, i.e., caching the input or output  datas at the mobile VR device, requires careful design. 
Such system model can also be commonly seen in other communication-intensive, computation-intensive and delay-sensitive applications, such as online gaming and augment reality (AR) \cite{E}.


%

\textbf{\textit{Contribution.}} In this paper, we  aim at jointly utilizing the computing and caching resources at mobile devices \cite{3C} and content-centric multicast \cite{multicast} to minimize the bandwidth requirement under the latency constraint for quality of experience (QoE). In particular,  we first show that the joint caching and computing optimization problem is NP-hard in strong sense, and then transform it into an equivalent difference of convex (DC) problem, which allows us to use concave convex procedure (CCCP) to obtain a stationary point. Moreover, in the symmetric scenario, i.e., $I_f=I$, $w_f = w$, $O_f=O$, $C_k=C$, $f_k=f_1$ and $\bar{E}_k=\bar{E}$ for all $f\in\mathcal{F}$ and $k\in \mathcal{K}$, we address the following two questions:

1) \textbf{Compared with MEC computing, how much gain on the bandwidth requirement on earth can the caching and computing resources at mobile devices bring?}
We find that when the size of output data is smaller than that of input data, the ratio of the minimum bandwidth requirement $B_{MEC}^*$ of MEC computing to that of the proposed system $B^*$ is
\begin{equation}
\frac{B_{MEC}^*}{B^*} = \frac{F}{F-\frac{C}{O}} \nonumber,
\end{equation}
implying that only caching at mobile devices can bring gain. Otherwise, the gain depends on both local caching and computing, e.g.,
\begin{equation}
\frac{B_{MEC}^*}{B^*} = \frac{F}{F-\frac{C}{O}-(\alpha -1) \frac{F\bar{E}}{\mu Owf_1^2}},\nonumber
\end{equation}
when $f_1 \geq \sqrt{\frac{F\bar{E}}{\mu wC}}$. Here,  $\alpha \triangleq \frac{O}{I}$  and $\mu$ is a constant related to the hardware architecture.

2) \textbf{Compared with unicast transmission, how much gain on bandwidth can the multicast transmission bring?}
We further show that the ratio of the minimum bandwidth requirement $B_{unicast}^*$ of unicast transmission to that of multicast transmission $B^*$ is
\begin{equation}
\frac{B_{unicast}^*}{B^*} = \frac{K}{F(1-(1-\frac{1}{F})^K)},\nonumber
\end{equation}
implying that the gain only depends on the number of mobile devices and that of tasks.

\textbf{\textit{Related Works.}} Our considered setting looks very similar to that in \cite{Ali} which studies the fundamental limits of caching, but it is very different. The core idea of \cite{Ali} is how to design the cache placement and coded delivery scheme to achieve global caching gain, which does not exploit the computing resources at mobile devices. As shown in \cite{3C}, computing is one of the three primary resources of mobile systems, and thus this paper is not a simple extension of \cite{Ali}. On the other hand, the proposed system is also different from traditional MEC systems \cite{mao}. In the traditional MEC system, mobile devices offload tasks to the MEC server to reduce latency or local energy consumption, and then the MEC delivers the computation results to mobile devices after computation. However, this approach generally increases the communication resource consumption, and thus it may be not always suitable for high bandwidth consumption application, e.g., VR video streaming illustrated in the motivating example. In addition, taking advantage of computation to reduce the communication load is studied in \cite{song} by using coded caching method \cite{Ali} in a distributed computing system. However, the computation load in \cite{song} is defined as the average number of nodes to compute one function, similar to the caching concept in \cite{Ali}.

Joint caching and computing at mobile devices has been studied in \cite{Yang} and our previous work \cite{Sunvr}. \cite{Yang} exploits the caching and computing resources at the mobile device to minimize the traffic load over wireless link. \cite{Sunvr} obtains the closed-form expression of the minimum average transmission rate, and analytically illustrates the tradeoff among communication, computing and caching. Note that \cite{Yang} and \cite{Sunvr} only consider a single-user setting and can not reveal the content-centric multicast gain for the bandwidth requirement.
\section{System Model}
As illustrated in Fig.~\ref{model}, we consider a multi-user MEC system consisting of one single-antenna MEC server and $K$ single-antenna mobile devices. The system operates over an infinite time horizon and time is slotted, indexed by $t=0,1,2,\cdots$. At the beginning of every time slot, each mobile device has a computation-intensive and communication-intensive task to be completed under latency constraint $\tau$ (\emph{in seconds}).\footnote{The duration of each time slot is assumed to be $\tau$ seconds.} 
The MEC server serves the mobile devices via content-centric multicast transmission. In particular, the  
 MEC server has access to the input and output datas of all the tasks.  Each mobile device is endowed with finite caching and computing resources, which can be utilized to reduce bandwidth requirement of the wireless multicast channel.

\subsection{Request Model}

We consider the applications the input datas of which are not generated from mobile devices and require downloading from the MEC server.  The task request stream at each mobile device is assumed to conform to independent reference model (IRM) based on the following assumptions: i) the tasks that each mobile device $k$ requests are fixed to the set $\mathcal{F}$; ii) the probability of the request for task $f$ at mobile device $k$, denoted as $P_{k,f}$, is constant and independent of all the past requests. We have $\sum_{f\in \mathcal{F}} P_{k,f} = 1$, for all $k\in \mathcal{K}$. Denote with $A_k \in \mathcal{F}$ the task requested by mobile device $k$, and $\mathbf{A} \triangleq (A_k)_{k\in \mathcal{K}} \in \mathcal{F}^K$ the system task request state, where $\mathcal{F}^K$ represents the system task request space. We assume that the $K$ task request processes are independent of each other, and thus we have $P(\textbf{A}) = \prod_{k\in \mathcal{K}} P_{k,A_k}$. 

In addition, we assume that each task request must be satisfied within a given time deadline of $\tau$ seconds for QoE. For example, in VR video streaming, $\tau\! \approx\! 20$\!~ms to avoid dizziness and nausea \cite{E}.

\subsection{Caching and Computing Model}
First, consider the cache placement at mobile device $k$, for all $k\in \mathcal{K}$. We assume that each mobile device $k$ is equipped with a cache size $C_k$ (\emph{in bits}), and is able to store both input and output datas of some tasks.
Denote with $c_{k,f}^I \in \{0,1\}$ the caching decision for input data of task $f$, where $c_{k,f}^I = 1$ means that the input data of task $f$ is cached in the mobile device $k$, and $c_{k,f}^I = 0$ otherwise. Denote with $c_{k,f}^O \in \{0,1\}$ the caching decision for output data of task $f$, where $c_{k,f}^O = 1$ means that the output data of task $f$ is cached in the mobile device $k$, and $c_{k,f}^O = 0$ otherwise. 
Under the cache size constraint, we have
\begin{equation}\label{CacheSize}
\sum_{f=1}^F I_fc_{k,f}^I + O_f c_{k,f}^O \leq C_k,\ k\in \mathcal{K}.
\end{equation}
Denote with $(\textbf{c}^I,\textbf{c}^O)$ the system caching decision, where $\textbf{c}^I \triangleq (c_{k,f}^I)_{k\in \mathcal{K}, f\in \mathcal{F}}$ and $\textbf{c}^O \triangleq (c_{k,f}^O)_{k\in \mathcal{K}, f\in \mathcal{F}}$ satisfy the cache size constraint in (\ref{CacheSize}).

Next, consider the computing decision at mobile device $k$, for all $k\in \mathcal{K}$. We assume that each mobile device $k$ is equipped with a computing server, which can run at a constant CPU-cycle frequency $f_k$ (\emph{in cycles/s}) and with a fixed average energy $\bar{E}_k$ (\emph{in J}). The power consumed at the mobile device for computation per cycle  with frequency $f_k$ is $\mu f_k^3$. Denote with $d_{k,f} \in \{0,1\}$ the computation decision for task $f$, where $d_{k,f}= 1$ means that task $f$ is computed at the mobile device $k$, and $d_{k,f} = 0$ otherwise. Under the average energy consumption constraint, we have
\begin{equation}\label{energy}
\sum_{f=1}^F P_{k,f}\mu f_k^2 I_fw_f d_{k,f} \leq \bar{E}_k,\ k\in \mathcal{K}.   
\end{equation}
Denote with $\textbf{d} \triangleq (d_{k,f})_{k\in \mathcal{K},f\in \mathcal{F}}$ the system computing decision, which satisfies the average energy consumption constraint in (\ref{energy}). 

\subsection{Service Mechanism}

Based on the joint caching and computing decision, i.e., $(\textbf{c}^I,\textbf{c}^O,\textbf{d})$, we can see that request for task $f \in \mathcal{F}$ at mobile device $k \in \mathcal{K}$ can be served via the following four routes, each of which yields a unique transmission rate requirement. Denote with $R_{f,j}^k$ (\emph{in bits/s}) the minimum transmission rate required for satisfying task $f$ at mobile device $k$ via Route~$j \in \{1,2,3,4\}$ within the deadline $\tau$ seconds.
 
\begin{itemize}
\item \textit{\textbf{Route~1: Local output caching}}. If $c_{k,f}^O=1$, i.e., the output data of task $f$ has been cached at the mobile device $k$, request for task $f$ can be satisfied directly from the cache of mobile device $k$, thereby without any need of computing or transmission. Thus, the required latency is negligible and $R_{f,1}^k = 0$.
\item \textit{\textbf{Route~2: Local input caching with local computing}}. If $c_{k,f}^O=0$, but $c_{k,f}^I=1$ and $d_{k,f} = 1$, i.e., the input data of task $f$ has been cached and task $f$ is chosen to be computed at the mobile device $k$, request for task $f$ can be satisfied via local computing based on the cached input data, thereby without any need of transmission. Thus, the required latency is $\frac{I_fw_f}{f_k}$ and $R_{f,2}^k = 0$. For feasibility, we assume that $\frac{I_fw_f}{f_k} \leq \tau, f\in \mathcal{F}, k\in \mathcal{K}$. 
\item \textit{\textbf{Route~3: Local computing}}. If $c_{k,f}^O=0$, $c_{k,f}^I=0$ and $d_{k,f} = 1$, i.e., the output or input data of task $f$ has not been cached and task $f$ is chosen to be computed at the mobile device $k$, the execution for satisfying task $f$ consists of the following two stages: i) the input data of task $f$ is transmitted from the MEC server; ii) the input data is computed at the mobile device $k$. Thus, the required latency is $\frac{I_f}{R_{f,3}^k} + \frac{I_fw_f}{f_k}$. Under the latency constraint, we have $\frac{I_f}{R_{f,3}^k} + \frac{I_fw_f}{f_k} = \tau$, i.e., $R_{f,3}^k = \frac{I_f}{\tau-\frac{I_fw_f}{f_k}}$. 
\item \textit{\textbf{Route~4: MEC computing}}. If $c_{k,f}^O = 0$, $c_{k,f}^I = 0$ and $d_{k,f} = 0$, i.e., output or input data of task $f$ has not been cached and task $f$ is not chosen to be computed locally, task $f$ is satisfied via downloading the output data from the MEC server. Thus, the required latency is $\frac{O_f}{R_{f,4}^k}$. Under latency constraint, we have $\frac{O_f}{R_{f,4}^k} = \tau$, i.e., $R_{f,4}^k = \frac{O_f}{\tau}$. 
\end{itemize}

In summary, denote with $x_{f,j}^k\in \{0,1\}$ the service decision for task $f$ at mobile device $k$, where $x_{f,j}^k = 1$ means that task $f$ at mobile device $k$ is served via Route~$j\in \{1,2,3,4\}$, and $x_{f,j}^k =0$ otherwise. To guarantee that task $f$ at mobile device $k$ gets served, we have 
\begin{equation}\label{schedule}
\sum_{j=1}^4 x_{f,j}^k = 1,\ f\in \mathcal{F},\ k\in \mathcal{K}. 
\end{equation}

In addition, the cache size and average energy consumption constraints in (\ref{CacheSize}) and (\ref{energy}) can be rewritten as 
\begin{equation}\label{cachesize2}
\sum_{f=1}^F I_f x_{f,2}^k + O_f x_{f,1}^k \leq C_k,\ k\in \mathcal{K}, 
\end{equation}
\begin{equation}\label{energy2}
\sum_{f=1}^F P_{k,f}\mu f_k^2I_fw_f(x_{f,2}^k+x_{f,3}^k) \leq \bar{E}_k,\ k\in \mathcal{K}. 
\end{equation}

For clarity, we illustrate the relationship between the service policy $\textbf{x} \triangleq (x_{f,j}^k)_{f\in \mathcal{F}, j\in \{1,2,3,4\},k\in \mathcal{K}}$ and joint caching and computing policy, i.e., $(\textbf{c}^I,\textbf{c}^O,\textbf{d})$, and the tradeoff among caching, computing and communication in Table~I. 
\begin{table}[t]
\caption{Tradeoff among Communication, Computing and Caching}\label{tradeoff}
\newcommand{\tabincell}[2]{\begin{tabular}{@{}#1@{}}#2\end{tabular}}
\begin{center}
\begin{tabular}{lccc}
\hline
\ \ \ \ \ \ \ \ \ \ Service\! Route  & Rate  & Caching\! Cost & Computing\! Cost \\
\hline
\tabincell{c}{$x_{f,1}^k\!=\!1$\\($c_{k,f}^O\!=\!1,\!c_{k,f}^I\!=0,\! d_{k,f}\!=\!0$)}  & $0$ & $O_f$ & $0$\\
\hline
\tabincell{c}{$x_{f,2}^k\!=\!1$\\($c_{k,f}^O\!=\!0,\!c_{k,f}^I\!=1, \!d_{k,f}\!=\!1$)} & $0$ & $I_f$  & $P_{k,f}\mu I_fw_ff_k^2$\\
\hline
\tabincell{c}{$x_{f,3}^k\!=\!1$\\($c_{k,f}^O\!=\!0,\!c_{k,f}^I\!=0, \!d_{k,f}\!=\!1$)}   & $R_{f,3}^k$ & $0$ & $P_{k,f}\mu I_fw_ff_k^2$ \\ 
\hline
\tabincell{c}{$x_{f,4}^k\!=\!1$\\($c_{k,f}^O\!=\!0,\!c_{k,f}^I\!=0, \!d_{k,f}\!=\!0$)}  & $R_{f,4}^k$ & $0$ & $0$\\
\hline
\end{tabular}
\end{center}
\label{a}
\end{table}%

\subsection{Multicast Transmission Model}
At each time slot, given system task request state $\textbf{A}$ and service decision $\textbf{x}$, we consider that the MEC server employs content-centric multicast to simultaneously serve many different requests for either input or output data of the same task. 
Specifically, denote with $B_f^I(\textbf{x}, \textbf{A})$ and $B_f^O(\textbf{x},\textbf{A})$ (in \emph{Hz}) the bandwidth allocated by the MEC server for transmitting the input and output data of task $f\in \mathcal{F}$, respectively. To guarantee each user's QoE, we have 
\begin{align}\label{input}
&B_f^I(\textbf{x},\textbf{A})\min_{k\in \mathcal{K}} \log\left(1+\frac{Ph_k^2}{\sigma^2}\right)\textbf{1}(A_k=f)x_{f,3}^k\nonumber\\
&\ \ \ \ \ \ \ \ \ \ \ \ \  \ \ \ \ \ \ \ \ \geq \max_{k\in \mathcal{K}} R_{f,3}^k\textbf{1}(A_k=f)x_{f,3}^k,\ f\in \mathcal{F}, 
\end{align}
\begin{align}\label{output}
&B_f^O(\textbf{x},\textbf{A})\min_{k\in \mathcal{K}} \log\left(1+\frac{Ph_k^2}{\sigma^2}\right)\textbf{1}(A_k=f)x_{f,4}^k\nonumber\\
&\ \ \ \ \ \ \ \ \ \ \ \ \  \ \ \ \ \ \ \ \ \geq \max_{k\in \mathcal{K}} R_{f,4}^k\textbf{1}(A_k=f)x_{f,4}^k,\ f\in \mathcal{F}, 
\end{align}
where $P$ denotes the transmission power of the MEC server, $\sigma^2$ denotes the variance of complex white Gaussian channel noise, and $h_k$ denotes the channel gain for mobile device $k$, which is assumed to be constant within the deadline $\tau$ seconds, respectively. $\textbf{1}(\cdot)$ denotes the indicator function throughout the paper.

Under $\textbf{x}$, denote with $B(\textbf{x})$ the average bandwidth requirement, and we have 
\begin{equation}\label{averagerate}
B(\textbf{x}) = \mathbb{E}\left[ \left(\sum_{f=1}^F B^I_f(\textbf{x},\textbf{A}) + B^O_f(\textbf{x},\textbf{A})\right)\right],
\end{equation}
where the expectation is taken over system request state $\textbf{A}\! \in\! \mathcal{F}^K$. 


\section{Problem Formulation}
In this paper, our objective is to minimize the average bandwidth requirement subject to the latency,  cache size and average energy consumption constraints. The optimization problem can be formulated as the following $0$-$1$ integer-programming problem. 
\begin{Prob}[Average Bandwidth Minimization]\label{Prob1}
\begin{align}
& \min_{\textbf{x}} \ \ \ \ \ \ \ \ \ \ \ \ \ \ \ \ \ \ \ \ \ \ \ \ \ \ B(\textbf{x}) \nonumber\\
& \ s.t. \ \ \ \ \ \ \ \ \ \ \ \ \ \ \ \ \ \ \ \ (\ref{schedule}),(\ref{cachesize2}),(\ref{energy2}),(\ref{input}),(\ref{output}), \nonumber\\
&\ \ \ \ \ \ x_{f,j}^k \in \{0,1\},\ f\in \mathcal{F},\ k\in \mathcal{K},\ j\in \{1,2,3,4\}.\label{binary1}
\end{align}
\end{Prob}
Denote with $B^*$ the minimum average bandwidth, and $\textbf{x}^*$ the optimal service decision. Thus, we have $B^* = B(\textbf{x}^*)$. Based on $\textbf{x}^*$, the optimal joint caching and computing policy, denoted as $(\textbf{c}^{I*},\textbf{c}^{O*}, \textbf{d}^*)$, can be obtained directly according to Table~I. 

It is direct to observe that (\ref{input}) and (\ref{output}) are reduced to equality for optimality, i.e., 
\begin{align}\label{input1}
&B_f^I(\textbf{x},\textbf{A}) = \max_{k\in \mathcal{K}} \frac{1}{\log\left(1+\frac{Ph_k^2}{\sigma^2}\right)}\textbf{1}(A_k=f)x_{f,3}^k\nonumber\\
&\ \ \ \ \ \ \ \ \ \ \ \ \  \ \ \ \ \ \ \ \  *\max_{k\in \mathcal{K}} R_{f,3}^k\textbf{1}(A_k=f)x_{f,3}^k,\ f\in \mathcal{F}, 
\end{align}
\begin{align}\label{output1}
&B_f^O(\textbf{x},\textbf{A}) = \max_{k\in \mathcal{K}} \frac{1}{\log\left(1+\frac{Ph_k^2}{\sigma^2}\right)}\textbf{1}(A_k=f)x_{f,4}^k\nonumber\\
&\ \ \ \ \ \ \ \ \ \ \ \ \  \ \ \ \ \ \ \ \ *\max_{k\in \mathcal{K}} R_{f,4}^k\textbf{1}(A_k=f)x_{f,4}^k,\ f\in \mathcal{F}. 
\end{align}

In the following, we will show that Problem~\ref{Prob1} is NP-hard in strong sense in Lemma~\ref{np}.\footnote{We omit all the proofs due to page limitation. Please refer to \cite{chenzhiyong} for details. } Then, we formulate the problem as a DC problem, which allows us to use CCCP to obtain a stationary point of Problem~\ref{Prob1}. 

\begin{lemma}[Computation Intractability of Problem~\ref{Prob1}]\label{np}Problem~\ref{Prob1} is NP-hard in strong sense.
\end{lemma}

\subsection{Equivalent Formulation}
In order to solve Problem~\ref{Prob1}, we transform it into Problem~\ref{equimmkp} without loss of equivalence. 

First, for all $f\in \mathcal{F}$ and $\textbf{A} \in \mathcal{F}^K$, let us introduce auxiliary variables, i.e.,  $a^I(f,\textbf{A})$, $b^I(f,\textbf{A})$, $a^O(f,\textbf{A})$ and $b^O(f,\textbf{A})$ satisfying 
\begin{align}\label{ai}
&a^I(f,\textbf{A}) = \max_{k\in \mathcal{K}} \frac{1}{\log\left(1+\frac{Ph_k^2}{\sigma^2}\right)}\textbf{1}(A_k=f)x_{f,3}^k,\nonumber\\
&\hspace{5cm}\textbf{A} \in \mathcal{F}^K,\ f\in \mathcal{F}, 
\end{align}
\begin{align}\label{bi}
b^I(f,\textbf{A}) \!=\! \max_{k\in \mathcal{K}} R_{f,3}^k\!\textbf{1}(A_k=f) x_{f,3}^k,\ \textbf{A} \in \mathcal{F}^K, f \in \mathcal{F}, 
\end{align}
\begin{align}\label{ao}
&a^O(f,\textbf{A}) = \max_{k\in \mathcal{K}} \frac{1}{\log\left(1+\frac{Ph_k^2}{\sigma^2}\right)}\textbf{1}(A_k=f)x_{f,4}^k,\nonumber\\
&\hspace{5cm}\textbf{A} \in \mathcal{F}^K,\ f\in \mathcal{F}, 
\end{align}
\begin{align}\label{bo}
b^O(f,\textbf{A}\!)\! =\! \max_{k\in \mathcal{K}} R_{f,4}^k\!\textbf{1}(A_k=f)x_{f,4}^k, \ \textbf{A} \in \mathcal{F}^K, f \in \mathcal{F}, 
\end{align}
respectively. 
 Accordingly, $B^I_f(\textbf{x},\textbf{A})$ and $B^O_f(\textbf{x},\textbf{A})$ can be rewritten as 
\begin{equation}\label{input2}
B_f^I(\textbf{x},\textbf{A})\! =\! \frac{\left(a^I(f,\textbf{A}) \!+ b^I(f,\textbf{A})\right)^2}{4}\!- \frac{\left(a^I(f,\textbf{A}) \!- b^I(f,\textbf{A})\right)^2}{4}, 
\end{equation}
\begin{equation}\label{output2}
B_f^O(\textbf{x},\!\textbf{A})\! =\! \frac{\left(a^O(f,\textbf{A}) \!+\! b^O(f,\textbf{A})\right)^2}{4}\!-\! \frac{\left(a^O(f,\textbf{A})\! -\! b^O(f,\textbf{A})\right)^2}{4}, 
\end{equation}
respectively. 

Secondly, note that (\ref{binary1}) can be rewritten as 
\begin{equation}\label{aa}
x_{f,j}^k \in [0,1], \ f \in \mathcal{F},\ j \in \{1,2,3,4\},\ k\in \mathcal{K},
\end{equation}
\begin{equation}\label{b}
\sum_{k=1}^K\sum_{f=1}^F \sum_{j=1}^4 x_{f,j}^k(1-x_{f,j}^k) \leq 0. 
\end{equation}

Then, by substituting (\ref{binary1}) with (\ref{aa}) and (\ref{b}), $B_f^I(\textbf{x},\textbf{A})$ and $B_f^O(\textbf{x},\!\textbf{A})$ with (\ref{input2}) and (\ref{output2}), respectively, and adding constraints (\ref{ai})-(\ref{bo}), we transform Problem~\ref{Prob1} into Problem~\ref{equimmkp} equivalently. 
\begin{Prob}[Equivalent Optimization]\label{equimmkp}
\begin{align}
& \min_{\textbf{x}} \ \ \ \ \ \ \  \ \ \ \ \ \ \ \ \ \ \ \ \ \ \ \ \ \ \ B(\textbf{x})  \nonumber \\
& \ s.t.  (\ref{schedule}), (\ref{cachesize2}), (\ref{energy2}),(\ref{ai}), (\ref{bi}), (\ref{ao}), (\ref{bo}),(\ref{input2}),(\ref{output2}),(\ref{aa}),(\ref{b}).  \nonumber
\end{align} 
\end{Prob}

Note that Problem~\ref{equimmkp} is a continuous optimization problem, the computation complexity of which is much less compared with solving Problem~\ref{Prob1} directly. However, considering \!$\sum_{k=1}^K\sum_{f=1}^F \!\sum_{j=1}^4\! x_{f,j}^k(1-x_{f,j}^k)$ in (\ref{b}) is a concave function, (\ref{b}) is not a convex constraint, and thus obtaining an efficient algorithm for solving Problem~\ref{equimmkp} is still very challenging. 



\subsection{Penalized Formulation and CCCP}
To facilitate the solution, we transform Problem~\ref{equimmkp} into Problem~\ref{penalized} by penalizing the concave constraints in (\ref{b}) to the objective function. 

\begin{Prob}[Penalized Optimization]\label{penalized}
\begin{align}
& \min_{\textbf{x}} \ \ \ \ \ \ \ \ \ \ \ \  \  B(\textbf{x})-\rho \sum_{k=1}^K\sum_{f=1}^F\sum_{j=1}^4x_{f,j}^k(x_{f,j}^k-1)\nonumber \\
& s.t. \ \  (\ref{schedule}), (\ref{cachesize2}), (\ref{energy2}),(\ref{ai}), (\ref{bi}), (\ref{ao}), (\ref{bo}), (\ref{input2}),(\ref{output2}),(\ref{aa}), \nonumber
\end{align} 
where the penalty parameter $\rho>0$. 
\end{Prob}

Note that the objective function of Problem~\ref{penalized} can be decomposed into a difference of two convex functions, and the constraints of Problem~\ref{penalized} are linear. Thus, Problem~\ref{penalized} is a DC problem. Based on Theorem~5 and Theorem~8 in \cite{exactpenalty}, we show the equivalence between Problem~\ref{equimmkp} and Problem~\ref{penalized} in the following lemma. 

\begin{lemma}[Exact Penalty]\label{exact} There exists $\rho_0>0$ such that when $\rho \geq  \rho_0$, Problem~\ref{penalized} and Problem~\ref{equimmkp} have the same optimal solution. 
\end{lemma}

Lemma~\ref{exact} illustrates that Problem~\ref{penalized} is equivalent to Problem~\ref{equimmkp} if the penalty parameter $\rho$ is sufficiently large. 
Thus, in the sequel, we solve Problem~\ref{penalized} instead of Problem~\ref{equimmkp} by using CCCP to obtain the stationary point \cite{exactpenalty}. In general, CCCP involves iteratively solving a sequence of convex problems, each of which is obtained via linearizing the concave-term of the objective function of Problem~\ref{penalized}. Specifically, based on CCCP,  at each iteration $t$,  we approximate $\frac{\left(a^I(f,\textbf{A}) - b^I(f,\textbf{A})\right)^2}{4}$ with $\frac{\left(a^{I}_t(f,\textbf{A}) - b^{I}_t(f,\textbf{A})\right)^2}{4} + \frac{a^{I}_t(f,\textbf{A})-b^{I}_t(f,\textbf{A})}{2}(a^{I}(f,\textbf{A})-b^{I}(f,\textbf{A})-(a^{I}_t(f,\textbf{A})-b^{I}_t(f,\textbf{A})))$,  $\frac{\left(a^O(f,\textbf{A}) - b^O(f,\textbf{A})\right)^2}{4}$ with $\frac{\left(a^{O}_t(f,\textbf{A}) - b^{O}_t(f,\textbf{A})\right)^2}{4} + \frac{a^{O}_t(f,\textbf{A})-b^{O}_t(f,\textbf{A})}{2}(a^{O}(f,\textbf{A})-b^{O}(f,\textbf{A})-(a^{O}_t(f,\textbf{A})-$$b^{O}_t(f,\textbf{A})))$ and $x_{f,j}^k(x_{f,j}^k\!-\!1)$ with $x_{f,j}^{k,(t)}(x_{f,j}^{k,(t)}\!-\!1)+ (2x_{f,j}^{k,(t)}-1)(x_{f,j}^k-x_{f,j}^{k,(t)})$. 
In order to obtain a global optima of Problem~\ref{equimmkp}, we obtain multiple local optimal solutions of Problem~\ref{penalized} via performing CCCP multiple times, each with a unique initial feasible point of Problem~\ref{penalized}, and then choose the one which achieves the minimum average value \cite{infeasible}. 

 \begin{figure}[t]
\begin{center}
 \includegraphics[width=7.8cm]{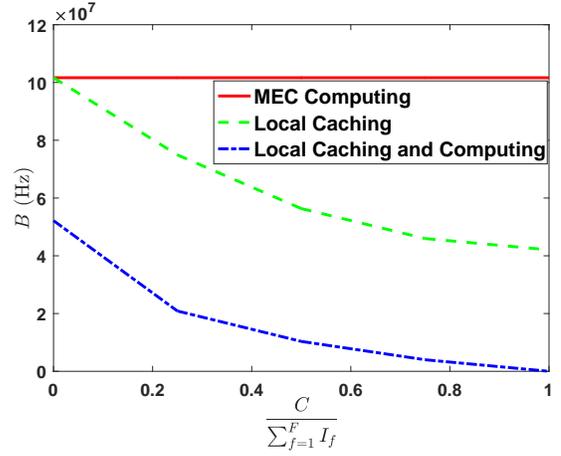}
\end{center}
 \caption{\small{Impact of $C$. $C_k=C,\ f_k=f_1$ and $\frac{1}{\log(1+\frac{Ph_k^2}{\sigma^2})} = 0.1$ for all $k\in \mathcal{K}$,  $F = 20$, $K = 2$, $I_f \in [1,15]$ M bits, $\alpha =2$, $w = 10\ cycles/bit$, $\mu=10^{-27}$, $f_1= 1.1*10^{11}Hz$, $\bar{E} = 1.7*10^3J $, $P_{k,f} \propto \frac{1}{f^\gamma}$ with $\gamma= 1$, $\rho = 10^4$.} }
\label{heteC}
\end{figure}
 \begin{figure}[t]
\begin{center}
 \includegraphics[width=7.8cm]{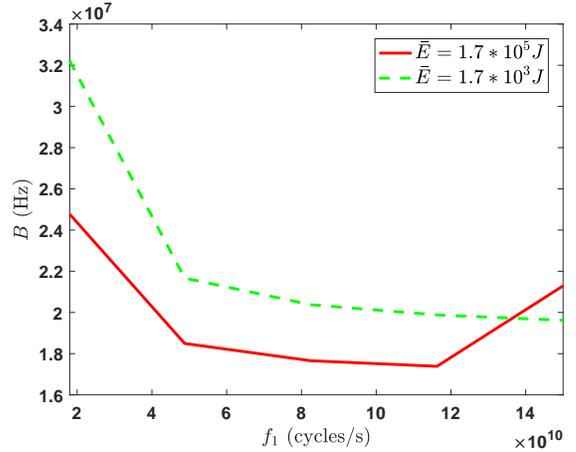}
\end{center}
 \caption{\small{Impact of $f_1$. $\frac{C}{\sum_{f=1}^FI_f} = 0.3$, and other parameters are the same as that in Fig.~\ref{heteC}.} }
\label{hetef1}
\end{figure}
We illustrate the impacts of $C$ and $f_1$ on the bandwidth $B^*$ in Fig.~\ref{heteC} and Fig.~\ref{hetef1}, respectively. In Fig.~\ref{heteC}, MEC computing represents the policy that each task request is served via MEC computing; local caching represents the policy that only caching resource at each mobile device is exploited via greedy algorithm; local caching and computing represents the policy obtained via CCCP.  From Fig.~\ref{heteC}, we can see that bandwidth decreases with $C$, and when $C=0$, there still exists bandwidth gain due to the local computing. In Fig.~\ref{hetef1}, $B$ is obtained via CCCP. We can see that when $\bar{E}$ is large enough, $B$ decreases with $f_1$ since increasing $f_1$ decreases bandwidth requirement via local computing; when $\bar{E}$ is limited, $B$  increases with $f_1$, since increasing $f_1$ decreases the number of tasks that can be computed locally.

\section{3C Tradeoff Analysis}
In this section, in order to obtain analytical results and provide more design insights, we consider the symmetric scenario, i.e., for all $f\in \mathcal{F}, \ k\in \mathcal{K}$, $(I_f,w_f,O_f) = (I,w,O)$, $P_{k,f} = \frac{1}{F}$, $f_k=f_1$, $C_k=C$, $\bar{E}_k = \bar{E}$ and $h_k=h$. Accordingly, we have $R_{f,3}^k = R_3$ and $R_{f,4}^k = R_4$, where $R_3 \triangleq \frac{I}{\tau-\frac{Iw}{f_1}}$ and $R_4 \triangleq \frac{O}{\tau}$, for all $f\in \mathcal{F}$ and $k\in \mathcal{K}$. For interest of design, we assume that $C \leq OF$ and $\frac{F\bar{E}}{\mu Iwf_1^2} \leq F$. 
\subsection{Optimal Policy}
First, by analyzing the structure of the problem, we obtain the optimal policy in the symmetric scenario, given as below.
\begin{lemma}[Optimal policy in symmetric scenario]\label{symmetric}
For all $k\in \mathcal{K}$,
\begin{equation}\label{optB1}
x_{f,1}^{k,*} = 
\begin{cases}
1,& f = 1,\cdots,n_1, \\
0,& \text{otherwise},
\end{cases}
\end{equation}
where $n_1 \triangleq \max\left\{\frac{C-\min\left\{C,\frac{F\bar{E}}{\mu w f_1^2}\right\}\textbf{1}(\alpha>1)}{O},0\right\}$,  
\begin{equation}
x_{f,2}^{k,*} = 
\begin{cases}
1,& f=n_1+1,\cdots,n_1+n_2,\\
0,&\text{otherwise},
\end{cases}
\end{equation}
where $n_2 \triangleq \min \left\{\frac{C}{I},\frac{F\bar{E}}{\mu Iwf_1^2}\right\}\textbf{1}(\alpha>1)$, 
\begin{equation}
x_{f,3}^{k,*} = 
\begin{cases}
1,& f=n_1+n_2+1,\cdots,n_1+n_2+n_3,\\
0,&\text{otherwise},
\end{cases}
\end{equation}
where $n_3\! \triangleq\! \left(\frac{F\bar{E}}{\mu Iwf_1^2}\!-\! \min \left\{\frac{C}{I}\!,\!\frac{F\bar{E}}{\mu Iwf_1^2}\right\}\right)\!\textbf{1}\left(\alpha\!>1, f_1\!>\!\frac{Iw}{(1\!-\!\frac{1}{\alpha})\!\tau}\!\right)$,
\begin{equation}
x_{f,4}^{k,*} = 
\begin{cases}
1,&f = n_1+n_2+n_3+1,\cdots,F,\\
0,& \text{otherwise}. 
\end{cases}
\end{equation}
\end{lemma}
From Lemma~\ref{symmetric}, note that when $\alpha \leq 1$, $x_{f,2}^{k,*}=x_{f,3}^{k,*}=0$ for all $k\in \mathcal{K}$ and $f\in \mathcal{F}$, meaning that joint local input caching and computing does not bring any bandwidth gain, and  the caching resources at all the mobile devices are utilized merely for output caching.
\subsection{Local Caching and Computing Gain}
Next, we analytically quantify the gain on the bandwidth requirement that caching and computing resources at the mobile devices can bring over MEC computing, i.e., the outputs of all the tasks are transmitted from the MEC server. Denote with $B_{MEC}^*$ the minimum bandwidth requirement via MEC computing. Based on Lemma~\ref{symmetric}, we obtain the following lemma.
\begin{theorem}[Local Caching and Computing Gain]\label{MECgain}
When $\alpha \leq 1$, 
\begin{equation}
\frac{B_{MEC}^*}{B^*} = \frac{F}{F-\frac{C}{O}},
\end{equation}
which increases with $C$ but is independent of $f_1$; when $\alpha > 1$, if $f_1 \geq \sqrt{\frac{F\bar{E}}{\mu wC}}$, 
\begin{equation}
\frac{B_{MEC}^*}{B^*} = \frac{F}{F-\frac{C}{O}-(\alpha -1) \frac{F\bar{E}}{\mu Owf_1^2}},
\end{equation}
which increases with $C$ and decreases with $f_1$; when $\alpha > 1$, if $\frac{Iw}{(1-\frac{1}{\alpha})\tau} < f_1 \leq \sqrt{\frac{F\bar{E}}{\mu wC}}$, 
\begin{equation}
\frac{B_{MEC}^*}{B^*} = \frac{F}{F-\frac{F\bar{E}}{\mu Iwf_1^2} + \frac{\tau}{\alpha(\tau-\frac{Iw}{f_1})} \left(\frac{F\bar{E}}{\mu Iwf_1^2}-\frac{C}{I}\right)},
\end{equation}
which increases with $C$ and $f_1$; when $\alpha > 1$, if  $f_1 \leq \frac{Iw}{(1-\frac{1}{\alpha})\tau}$, 
\begin{equation}
\frac{B_{MEC}^*}{B^*} = \frac{F}{F-\frac{C}{I}},
\end{equation}
which increases with $C$ and is independent of $f_1$. 
\end{theorem}
 \begin{figure}[t]
\begin{center}
 \includegraphics[width=7.5cm]{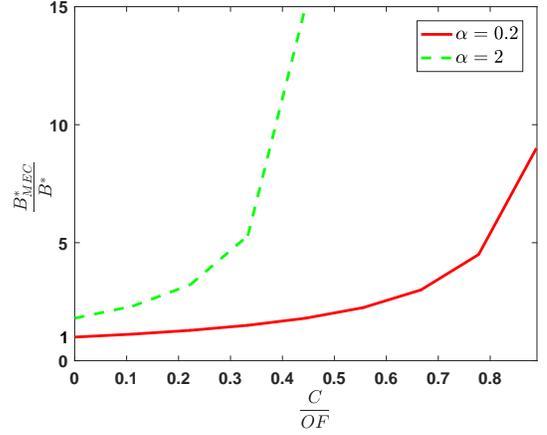}
\end{center}
 \caption{\small{Impact of $C$ on local caching and computing gain. $F = 50$, $K = 10$, $I = 15  M\ bits$, $w = 10\ cycles/bit$, $\mu=10^{-27}$, $f_1= 1.1*10^{11}Hz$, $\bar{E} = 1.7*10^3J $. }
 }
\label{multigainC}
\end{figure}

\begin{figure}[t]
\begin{center}
 \includegraphics[width=7.5cm]{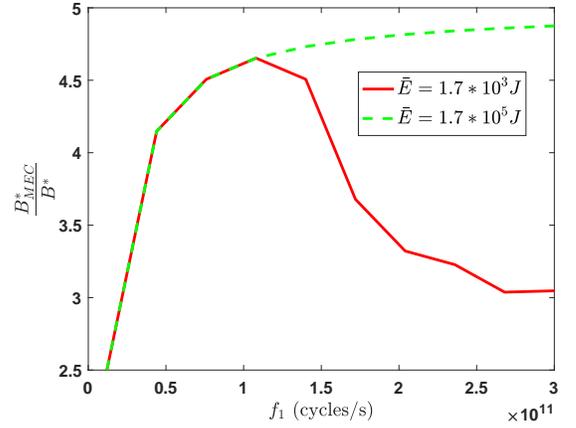}
\end{center}
 \caption{\small{Impact of $f_1$ on local caching and computing gain. $\frac{C}{OF} = 0.3$ and other parameters are the same as that in Fig.~\ref{multigainC}.}
 }
\label{multigainf1}
\end{figure}
Theorem~\ref{MECgain} shows that in the symmetric scenario, the local caching and computing gain depends on both the caching and computing capabilities of the mobile device, as illustrated in Fig.~\ref{multigainC} and Fig.~\ref{multigainf1}. Fig.~\ref{multigainC} shows that the local caching and computing gain increases with $C$, especially in the case when $\alpha >1$. That is, when $\alpha>1$, a small increase of $C$ incurs a significant increase of the gain.  From Fig.~\ref{multigainf1}, we can see that the local caching and computing gain increases with $f_1$ when the average energy is large enough, since increasing $f_1$ decreases the bandwidth requirement via local computing. However, the gain first increases and then decreases with $f_1 $ when the average energy is relatively limited, since increasing $f_1$ decreases the number of tasks that can be computed locally.

\subsection{Multicast Gain}
Finally, we analytically quantify the gain on the bandwidth requirement that multicast transmission can bring over unicast transmission, in which the MEC server transmits the requested datas by the mobile devices via $K$ independent unicast channels. The average bandwidth requirement for unicast transmission under $\textbf{x}$, denoted as $B_{unicast}(\textbf{x})$, is given by
\begin{equation}\label{unicast}
B_{unicast}(\textbf{x}) \!\triangleq\! \sum_{k=1}^K \sum_{f=1}^FP_{k,f}\!\sum_{j=1}^4\! R_{f,j}^k\frac{1}{\log(1+\frac{Ph_k^2}{\sigma^2})}x_{f,j}^k,  
\end{equation}
and denote with $B^*_{unicast}$ the minimum required bandwidth for unicast transmission. Based on Lemma~\ref{symmetric}, we obtain the multicast gain, i.e.,  $\frac{B_{unicast}^*}{B^*}$, given as below.
\begin{theorem}[Multicast Gain]\label{gain1}
\begin{equation}
\frac{B_{unicast}^*}{B^*} = \frac{K}{F(1-(1-\frac{1}{F})^K)},
\end{equation}
which decreases with $\frac{F}{K}$.
\end{theorem}

\begin{figure}[t]
\begin{center}
 \includegraphics[width=8cm]{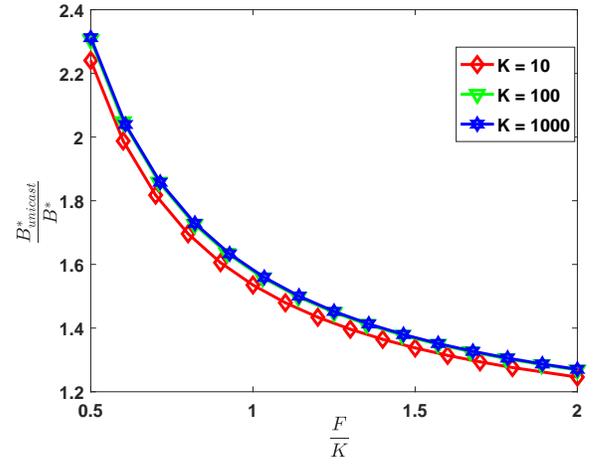}
\end{center}
 \caption{\small{Impact of $\frac{F}{K}$ on multicast gain.}
 }
\label{multigainK}
\end{figure}

Theorem~\ref{gain1} shows that in the symmetric scenario, the multicast gain depends only on the number of users $K$ and that of tasks $F$, as illustrated in Fig.~\ref{multigainK}. From Fig.~\ref{multigainK}, we can see that the multicast gain decreases with $\frac{F}{K}$. In addition,  multicast transmission achieves bandwidth gain only when $F \leq K$, and not otherwise. This is mainly because when $F>K$, the probability that multiple users request the same task decreases, and thus the multicast gain is negligible. 

\section{Conclusion}
In this paper, we investigate the impacts of the caching and computing resources at mobile devices on the transmission bandwidth, and optimize the joint caching and computing policy to minimize the average transmission bandwidth under the latency, local caching and local average energy consumption constraints. In particular, we first show the NP-hardness of the problem and transform it to a DC problem without loss of equivalence, which is solved efficiently via CCCP. In the symmetric scenario, we obtain the optimal joint policy and the closed form expressions for local caching and computing gain as well as multicast gain. In summary, we show theorectically that: in the symmetric scenario, 
\begin{itemize}
  \item $\frac{B_{MEC}^*}{B^*}$ increases with $C$;
    \item $\frac{B_{MEC}^*}{B^*}$ decreases with $f_1$ when $\alpha > 1$ and $f_1 > \sqrt{\frac{F\bar{E}}{\mu wC}}$; 
        \item $\frac{B_{MEC}^*}{B^*}$ increases with $f_1$ when $\alpha > 1$, $\frac{Iw}{(1-\frac{1}{\alpha})\tau} < f_1 \leq \sqrt{\frac{F\bar{E}}{\mu wC}}$; 
  \item $\frac{B_{MEC}^*}{B^*}$ remains unchanged with $f_1$ when $\alpha \leq 1$ or when $\alpha > 1$ and $f_1 \leq \frac{Iw}{(1-\frac{1}{\alpha})\tau}$;

 \item $\frac{B_{unicast}^*}{B^*}$ decreases with $\frac{F}{K}$ and $\frac{B_{unicast}^*}{B^*}=1$ when $F=K$.
\end{itemize}

\end{document}